# Solid-like to Liquid-like Behavior of Cu Diffusion in Superionic $Cu_2X$ (X=S, Se): An Inelastic Neutron Scattering and Ab-Initio Molecular Dynamics Investigation

Sajan Kumar[1,2], M. K. Gupta[1,2$], Prabhatasree Goel[1], R. Mittal[1,2*], Olivier Delaire[3], A. Thamizhavel[4], S. Rols[5] and  S. L. Chaplot[1,2]

[1]*Solid State Physics Division, Bhabha Atomic Research Centre, Trombay, Mumbai 400085, India*
[2]*Homi Bhabha National Institute, Anushaktinagar, Mumbai 400094, India*
[3]*Department of Mechanical Engineering and Materials Science, Duke University, NC 27708, USA*
[4]*Tata Institute of Fundamental Research, Homi Bhabha Road, Colaba, Mumbai 400005, India*
[5]*Institut Laue-Langevin, BP 156, 38042 Grenoble Cedex 9, France*
Email: rmittal@barc.gov.in[*], mayankg@barc.gov.in[$]

**Abstract**

$Cu_2Se$ and $Cu_2S$ are excellent model systems of superionic conductors with large diffusion coefficients that have been reported to exhibit different solid- liquid-like Cu-ion diffusion. In this paper, we clarify the atomic dynamics of these compounds with temperature-dependent ab-initio molecular dynamics (AIMD) simulations and inelastic neutron scattering (INS) experiments. Using the dynamical structure factor and Van-Hove correlation function, we interrogate the jump-time, hopping length distribution and associated diffusion coefficients. In cubic-$Cu_2Se$ at 500 K, we find solid-like diffusion with Cu-jump lengths matching well the first-neighbour Cu-Cu distance of ~3 Å in the crystal, and clearly defined optic phonons involving Cu-vibrations. Above 700 K, the jump-length distribution becomes a broad maximum cantered around 4 Å, spanning the first and second neighbour lattice distances, and a concurrent broadening of the Cu-phonon density of states. Further, above 900 K, the Cu-diffusion becomes close to liquid-like, with distributions of Cu-atoms continuously connecting crystal sites, while the vibrational modes involving Cu motions are highly damped, though still not fully over-damped as in a liquid. At low temperatures, the solid-like diffusion is consistent with previous X-ray diffraction and quasielastic neutron scattering experiments, while the higher-temperature observation of the liquid-like diffusion is in agreement with previous AIMD simulations. We also report AIMD simulations in $Cu_2S$ in the hexagonal and cubic superionic phases, and observe similar solid and liquid-like diffusion at low- and high-temperatures, respectively. The calculated ionic-conductivity is in fair agreement with reported experimental values.

# I. INTRODUCTION

In recent years, the world has seen an increased demand for green energy, which has the potential to mitigate the threat of global warming. Superionic materials play an important role in the domain of energy conversion and storage[1-7]. Copper chalcogenides ($Cu_2X$, X= Se, S) are superionic compounds with excellent thermoelectric properties[8, 9], attributed in part from the melting behavior of the Cu-sublattice in $Cu_2X$ (X=S, Se) at elevated temperatures, which results in an intrinsically low thermal conductivity[10-12]. Besides their potential as thermoelectric, these compounds are candidates for photovoltaics and photo electrodes owing to their favorable electrical, and optical properties[8, 13-17]. These materials show a rich phase diagram with temperature and also with Cu off-stoichiometry[18, 19], including high-temperature cubic phases that exhibit fast Cu ion diffusion[20, 21]. The added advantages of these compounds are that they are cheap, safer, abundant, and non-toxic, but the limitation is that high temperature is required for operation[22]. Typically, these two properties are required for a compound to exhibit superionicity: availability of a large number of vacant sites and low activation energy, so that ions can easily hop between available sites.

Copper selenide, $Cu_2Se$ crystallizes in a monoclinic phase (space group *C2/c*, β-phase) at ambient conditions, while above 414 K, it transforms to a superionic cubic phase (space group *Fm-3m*, α-phase)[22-24]. The β-phase of $Cu_2Se$ shows an ordered structure and exhibits low ionic conductivity[23]. In the high-temperature α-phase, the Cu diffusion coefficient is $\sim 10^{-4}$ to $10^{-5} cm^2 s^{-1}$ at 670 K and the corresponding ionic conductivity is 1-2 $\Omega^{-1} cm^{-1}$, which is three orders of magnitude higher than the *β*-phase at room temperature and similar to molten salts[25, 26]. The superionic phase can be stabilized to ambient temperature in Cu deficient $Cu_{2-x}Se$[27]. In the superionic α-phase at 415 K, the Se atoms occupy the 4*a* Wyckoff sites (FCC position in cubic phase), and Cu atoms partially occupy the 4*b* (octahedral), 8*c* (tetrahedral), and 32*f* Wyckoff sites (encloses the 8*c* tetrahedral sites) with fractional occupancies ~0.001, ~0.72, and ~0.07 respectively (**figure 1(a))**. The 32*f* sites form a tetrahedral cage around the 8*c* site[28-30].

The sister compound $Cu_2S$ crystallizes in an orthorhombic phase (*Pna2₁*) at temperatures below 202 K, and exhibits many polymorphs above this temperature: tetragonal (*P4₃2₁2*) above 202 K, hexagonal (*P6₃/mmc*, β-phase) above 370 K and cubic phases (*Fm-3m*, α-phase) above 700 K[31-35]. The hexagonal and cubic phases of $Cu_2S$ show superionic behavior[36]. In the superionic cubic α-phase, the S atoms occupy the 4*a* Wyckoff sites and Cu atoms partially occupy the 4*b*, 8*c*, and 192*l* Wyckoff sites with fractional occupancies ~0.09, ~0.26, ~0.03 respectively (**figure 1(b)**)[35]. The 192*l* sites form a cage around the 8*c* site. The details about the various structural phase transitions and range of stability of $Cu_2X$ (X=S, Se) are summarized in **Table SI**.

Superionic materials are often described as hybrid solid-liquid structures[8, 36-39], with the notion that at least one sub lattice has melted and shows significant diffusion. However, the molten sub lattice is usually not a simple liquid. For example, in the case of superionic Li$_2$O, Li occupies 8$c$ tetrahedral sites, and undergo[40, 41] jump diffusion from one lattice site to another, but with the jump time remaining much smaller than the residence time at either lattice site. Therefore, the time-averaged structure shows a high probability of the occurrence of the Li atoms at the sublattice sites, and therefore the crystal structure and its sites remain well defined in this case. As we shall discuss below, the diffusion of Cu in Cu$_2$X is more complex, in part because the Cu atoms can reside at many available sites in the crystal, as noted above.

Many investigations of Cu diffusion in Cu$_2$Se using quasielastic neutron scattering (QENS) have been reported[21, 23, 42]. These QENS results have generally been analyzed on the basis of jump-diffusion models assuming atomic jumps between fixed atomic sites in the crystal. However, the reported time and length scales of atomic jumps do not agree well with each other among these prior works. Ab-initio molecular dynamics (AIMD) simulations on Cu$_2$Se in Ref. [42] indicated a jump-diffusion behavior between the different tetrahedral (8$c$) and octahedral sites (4$b$) in the crystal. In contrast, another recent AIMD study [20] on Cu$_2$S and Cu$_2$Se proposed a liquid-like diffusion mechanism, with short residence times for Cu ions at the crystallographic sites. In the present study, we resolve the ambiguity in the nature of diffusion, i. e., whether it is solid-like jump diffusion between sub lattice sites, or, a liquid-like diffusion with very short residence times and low occupation probabilities of the Cu sub lattice crystallographic sites, by performing extensive AIMD simulations as a function of temperature. We find that, at moderate temperatures, the jump lengths show a broad distribution but that the Cu position probability distribution remains peaked at crystallographic sites, while the Cu-dynamics become more liquid-like at high temperatures. These results have important implication for the understanding of ionic conductivity, as well as the thermal conductivity in these materials. We also report inelastic neutron scattering measurements of the phonon spectra in Cu$_2$Se as a function of temperature, which sheds some light on the nature of diffusion and possible relationship of phonons and diffusion.

## II. EXPERIMENTAL DETAILS

We have prepared the Cu$_2$Se sample using solid-state reaction method and were characterized using X-ray diffraction at room temperature. All the peaks were well indexed with space group Fm-3m (**Figure S1** (Supplementary Material[43])). The refined Cu occupancies showed that the sample is slightly Cu deficient, with estimated composition of Cu$_{1.85}$Se. We used 10 grams of a polycrystalline sample of Cu$_{1.85}$Se to measure the phonon density of states (PDOS) in the temperature range from 100-450 K. The measurements were carried out using the time-of-flight IN4C spectrometer at the Institut Laue

Langevin(ILL), France. The spectrometer detector bank covers a wide range of scattering angles (10° to 110°). The measurements were conducted with 14.2 meV incident neutron energy in the energy gain mode. The neutron-weighted phonon density of states, $g^{(n)}(E)$ was obtained from the measured dynamic structure factor $S(Q, E)$ as given by[44-46]:

$$g^{(n)}(E) = A \left\langle \frac{e^{2W(Q)}}{Q^2} \frac{E}{n(E,T) + \frac{1}{2} \pm \frac{1}{2}} S(Q,E) \right\rangle \quad (1)$$

where A is a normalization constant and $Q$, $2W$, $E$ are the momentum transfer, Debye-Waller factor, and phonon energy. $n(E,T)$ is the Bose-Einstein population factor for phonon energy $E$ at temperature $T$. The +/- sign corresponds to energy loss/gain of the neutron. The total neutron-weighted density of states, $g^{(n)}(E)$ is related to the atomic partial density of states $g_k(E)$ as follows:

$$g^n(E) = B \sum_k \left\{ \frac{4\pi b_k^2}{m_k} \right\} g_k(E) \quad (2)$$

here B, $b_k$, and $m_k$ are a normalization factor, the neutron scattering length and mass of $k^{th}$ atom.

## III. COMPUTATIONAL DETAILS

We used density functional theory (DFT) for lattice and molecular dynamics simulations as implemented within the software package VASP (Vienna Ab-Initio Software Package)[47, 48]. The simulations were performed in the generalized gradient approximation (GGA) with the Perdew-Burke-Ernzerhof (PBE) exchange-correlation functional and projected augmented wave (PAW) pseudo-potentials with a kinetic energy plane-wave cutoff of 900 eV[49, 50].

The lattice dynamics calculations of $Cu_2Se$ and $Cu_2S$ were performed only in the ordered phase (i.e. monoclinic phase in $Cu_2Se$; orthorhombic and tetragonal phase in $Cu_2S$). For phonon calculations, we took the optimized structure and displaced symmetrically inequivalent atoms by 0.03Å in ±x, ±y, ±z directions and calculated the forces using Hellmann-Feynman theorem, which the PHONOPY software[51] then uses to compute the phonon dynamical matrix and dispersions. A 2×2×1 k-point mesh in the monoclinic $Cu_2Se$, and a 2 × 2 × 2 k-point mesh in orthorhombic and tetragonal $Cu_2S$ has been used. The k-points were generated using the Monkhorst-Pack method[52]. The force convergence criterion was set to $10^{-3}$ eV/Å², and the energy convergence criterion was set to $10^{-8}$ eV.

The ab-initio molecular dynamics simulations (AIMD) simulations were performed in NVT ensemble using the Nose-Hoover thermostat. The simulation was run for 60 ps with a 2 fs time steps. We equilibrate the system for ~10 picoseconds to attain the desired temperature. Simulations were executed for different phases with different temperatures (for Cu$_2$Se α-phase 500 K to 900K, β-phase 100K to 300 K; for Cu$_2$S α-phase 750K to 1100K, β-phase 300K to 700K, orthorhombic 100K, tetragonal 300K). The supercell dimensions used in AIMD simulations in different phases of Cu$_2$S and Cu$_2$Se phases are given in **Table SII** (Supplementary Material[43]). A single k-point sampling of the Brillouin zone at the zone center and an energy convergence of $10^{-6}$ eV were chosen.

The phonon density of states (PDOS) g(E) was calculated from AIMD simulations by taking the Fourier transform of the velocity autocorrelation function[53].

$$g(E) = \sum_{i=1}^{N} \frac{\int v_i(0).v_i(t) e^{iEt/\hbar} dt}{<v_i(0).v_i(0)>} \qquad (3)$$

where $v_i(t)$ is the $i^{th}$ atom velocity at time $t$ and $N$ is the total number of atom in supercell.

The diffusion coefficient $D$ has been estimated from three different techniques, namely, by using the mean-squared displacement (MSD), the velocity autocorrelation (VACF), and the dynamical structure factor $S(Q,E)$. We have calculated the time dependence of MSD using the following relationship[54, 55]

$$<u^2(\tau)> = \frac{1}{N_{ion}(N_{step}-N_\tau)} \sum_{i=1}^{N_{ion}} \sum_{j=1}^{N_{step}-N_\tau} |r_i(t_j + \tau) - r_i(t_j)|^2 \qquad (4)$$

where $r_i(t_j)$ is the $i^{th}$ atom position at $j^{th}$ time step, $N_\tau = \tau/\delta$, $\delta$ being the time-step in the simulation; $N_{step}$ is the total number of simulation steps, and $N_{ion}$ is the total number of a given element (Cu, S, Se) in a supercell. Using this time dependence, we estimated the diffusivity $D$ as given by[56]:

$$<D(T)> = <u^2(\tau)>/6\tau \qquad (5)$$

In addition, we also calculated $D$ by integrating the velocity autocorrelation function (VACF) as given by,

$$D = \int v_i(0).v_i(\tau) d\tau \qquad (6)$$

A few experimental and simulation investigations have discussed the Cu diffusion behavior in superionic phase of Cu$_2$Se; however, the values of $D$ estimated from different techniques (QENS, impedance spectroscopy, and simulations) shows significant variance ($10^{-5}$ to $10^{-7}$ cm$^2$s$^{-1}$)[20, 21, 23, 42].

Hence, we performed a detailed investigation of Cu diffusion using several approaches. Further, to address the ambiguity from QENS measurements, we have calculated the dynamical structure factor, $S(Q, E)$, using long AIMD trajectories of ~150 p.

Usually, the stochastic dynamics are analyzed by fitting the $S(Q, E)$ measured by QENS with one Lorentzian function and one delta function convoluted with the resolution function of the instrument. The $Q$-dependence of the half-width-at-half-maximum (HWHM) of the Lorentzian function provides information about the diffusion characteristics[57]. For localized diffusion, HWHM usually does not show significant variation with $Q$. While for long-range diffusion, HWHM in the low $Q$ regime is characterized by a $Q^2$ dependence. Further, in the case of solid-state diffusion, the $Q$-dependent HWHM data may be fitted using jump-diffusion models, such as the Chudley-Elliott (C-E) mode [58-60] to get the mean jump length and average residence time at a site. The C-E model assumes a fixed jump length, and the $Q$-dependent HWHM is given by:

$$\Gamma(Q) = \frac{\hbar}{\tau}(1 - \frac{Sin(Qd)}{(Qd)}) \quad (7)$$

where $\tau$ is the mean residence time, and $d$ is the jump length. And the diffusion constant, $D$ is given by
$$D = d^2/6\tau \quad (8)$$

## IV RESULTS AND DISCUSSIONS

### A. Temperature dependence of phonon spectra using INS and AIMD simulations

The temperature-dependent PDOS of $Cu_{1.95}Se$ was measured from 100 K to 450 K (**figure 2(a)**). This temperature range covers the monoclinic to cubic phase transition at 414 K. The energy range of the phonon spectrum extends up to 40 meV. We observe that at 100K, the INS spectra show sharp and well-defined peaks at 8 meV, 12 meV, and 22 meV as shown in **figure 2(a).** Our INS results are in good agreement with those Voneshen *et al*[21], although the energy resolution of the present data is superior due to the use of a smaller incident neutron energy and use of the neutron energy gain mode. Further, we have also calculated the partial contributions of Cu and Se to the PDOS and found that both the atoms contribute to the entire spectral range of vibrations **(figure 2(b))**. We note that the phonon dispersion relation of the acoustic phonons has been measured[30] in $Cu_{1.85}Se$ up to ~15 meV, which appears to be consistent with the present PDOS measurements. We have also performed lattice dynamics calculation in monoclinic phase of $Cu_2Se$. and low-T phases (tetragonal and orthorhombic phases) at T= 0K (**figure S2**). We found that the phonon spectra in the ordered low-T phases show sharp peaks in the PDOS.

At high temperatures, the peaks in the measured phonon spectrum get significantly broadened **(figure 2(a))**. The considerable broadening is attributed to large anharmonic lattice vibrations and Cu diffusion. To account for the anharmonic effects on the phonon spectrum, we performed AIMD calculations in the superionic cubic phase of $Cu_2X$ (X=S, Se). In **figure 3,** we show the total PDOS of $Cu_2X$ (X=S, Se) computed from AIMD in the superionic cubic phase at elevated temperatures. The PDOS does not show any sharp peak, and at elevated temperatures, the PDOS tends to a finite value at zero energy for both compounds, which is a signature of diffusion. Further, the PDOS value at $E$=0 increases with temperature, which also indicates a faster diffusion at higher temperatures. We also calculated the partial PDOS of different phases of $Cu_2X$ (X=S, Se) to see the contribution from different species to the PDOS and found that the finite PDOS at zero energy is mainly contributed from Cu dynamics (**figure S2 and S3 (**Supplementary Material[43]**))**. This further confirms that Cu is the only mobile ion at elevated temperatures, while S/Se atoms form a more rigid crystalline framework. Further, the phonon spectral weight shifts towards lower energies upon warming in both the compounds, which can be attributed to damping of strongly anharmonic phonon modes in the superionic regime. However, it is important to note that the phonon spectrum of the Cu atoms persists up to high energies in the superionic phase indicating that a fast vibrational behaviour of the Cu atoms remains despite the fast diffusion. This suggests that the nature of the diffusion continues to be jump-like up to moderate temperatures of ~700 K, with the atoms spending significant time undergoing vibrations around rest positions between jumps, as is characteristic in a solid. At higher temperatures ~900 K, the diffusion becomes closer to liquid-like. Therefore, we will discuss in the next section how the diffusion process is rather complex, involving both localized diffusion and long-ranged diffusion.

Besides the superionic cubic phase, we also computed the PDOS of $Cu_2S$ in the hexagonal superionic phase, which exists between 370 K to 700 K (**figure S3 (**Supplementary Material[43]**))**. Here also, a finite value of PDOS at $E$= 0 infers the presence of Cu diffusion; however, relatively less in magnitude than that in the superionic cubic phase.

**B. Cu Diffusion in $Cu_2X$ (X = S, Se ) and Van Hove Correlations $g_s(r, t)$**

In **figure 4**, we show the calculated MSD of Cu in $Cu_2S$ and $Cu_2Se$ at different temperatures. The Cu-MSD shows a nearly linear increase with time, characteristic of diffusion. However, the MSD values for Se/S oscillate about mean values consistent with a solid framework structure (**figure S4 (**Supplementary Material[43]**))**. To investigate the Cu-Cu and Cu-S/Se distances in the superionic cubic phase, we also calculated the time-averaged pair distribution function (PDF) from AIMD trajectories in superionic phases of $Cu_2X$ (X=S, Se) (**figure S5(**Supplementary Material[43]**))**. In both compounds, the first-neighbor peak for the Cu-Cu and Cu-Se PDF is quite sharp, while the peaks corresponding to farther neighbors are much broader and difficult to resolve.

We computed the self Van Hove correlation function, $g_s(r, t)$ of Cu in $Cu_2Se$ at 500 K, 700 K, and 900 K. The $g_s(r, t)$ computed at various time intervals from 2 ps to 26 ps are shown in **figure 5**. At all temperatures, $g_s(r, t)$ exhibits a peak around 1 Å in the first 2 ps, which gradually broadens with time; this implies that initially, Cu atoms are rattling within a cluster formed by 8$c$ and 32$f$ Wyckoff sites (the cluster diameter is ~ 1 Å). Further, $g_s(r, t)$ at 500 K, shows a peak around 3 Å at later times, attributed to the Cu jumps from one cluster to another cluster along various possible channels (see **Table I**). This maximum centered around 3 Å is mainly contributed from tetrahedral to tetrahedral (T-T) Cu hopping along (100) and tetrahedral to octahedral (T-O) hopping along (111). Although broad it retains a peak like structure in $g_s(r, t)$ at 500 K, suggesting that $Cu^+$ spends significant time near 4$b$/8$c$/32$f$ sites and hence the Cu sublattice retains a discrete set of sites and hence Cu diffusion mainly occurs through discrete jumps and could be well described by jump-diffusion models. Interestingly, at 700 K and 900 K, the first peak ~1 Å swiftly decays, and $g_s(r, t)$ develop a very broad distribution of about 4 Å, infers Cu spends less time in clusters around 8$c$ sites and quickly hops to the next cluster. At higher temperatures, it is difficult to define the peak in $g_s(r, t)$, and hence it seems the Cu sublattice melts in $Cu_2Se$ and behaves more like a liquid. The broad distribution around 4 Å contributed from all possible Cu diffusion channels, namely, (T-T) Cu hopping along with all possible directions ((100), (111) and (111)), and (T-O) and O-O hopping along (111). The gradual shift of $g_s(r, t)$ distribution at higher distances at elevated temperatures indicates a more favorable hopping of Cu along (110) directions. Further, the $g_s(r, t)$ at 500 K takes ~ 15-20 ps to span the distribution around 3 Å, while at 700 K within ~ 10 ps, it spans around 4 Å distance. This indicates that in $Cu_2Se$ at 500 K, the diffusion of Cu is solid-like however at high temperatures of 700 K and 900 K Cu diffusion can be described as liquid-like. Previous AIMD simulations[20, 42] at 700 K and 900 K also indicated liquid-like diffusion.

AIMD simulations of $Cu_2S$ at 900 K indicated[20] that diffusion of Cu is liquid-like, while the AIMD simulation on low temperature orthorhombic and hexagonal phases of $Cu_2S$ showed[36] that the hexagonal phase could be described as a solid-liquid hybrid phase. The time-average structure in a solid-liquid hybrid phase indicates that there is a high probability of the occurrence of the Cu atoms at the sublattice sites. The calculated self Van Hove correlation function, $g_s(r, t)$ in hexagonal and cubic phases of $Cu_2S$ are shown in **figure 5**. At 500 K, in the hexagonal $Cu_2S$, $g_s(r, t)$ shows a sharp peak around 1 Å, which slowly decays, and distribution around 2.5 Å develops with time. The peak distribution around 2.5 Å is contributed from Cu hopping between various Cu sites (6$g$-6$g$, 4$f$-4$f$, and 2$b$-6$g$ sites). At 700 K, the peak in $g_s(r, t)$ at ~ 1 Å becomes broad and swiftly decays, and very broad distribution of about 3 Å is evolved. Interestingly, in the cubic phase of $Cu_2S$, $g_s(r, t)$ at 750 and 900 K shows a much faster decay of 1 Å peak and shows a broad distribution around 4 Å, similar to the case of cubic $Cu_2Se$. Our analysis on $Cu_2S$ establishes that it also behaves like $Cu_2Se$, where at low temperature diffusion of Cu is solid like, while at 750 K and 900 K it behaves more like a liquid.

To visualize the diffusion process, we plotted the trajectories of a few Cu atoms in the superionic-cubic phase in $Cu_2Se$ at 700 K (**figure 6**). We can see that, at 700 K, Cu hops from one tetrahedral (8*c*) to another tetrahedral site (8*c*) either via an octahedral site (4*b*) or a direct jump, consistent with previous reports[23, 42]. Further, by observing the individual Cu trajectories, we could see that Cu ions rattle around 8*c* (i.e., hoping among 32*f* sites) for extended times before making a jump to the next tetrahedral site. The hopping among 32*f* sites is much faster and leads to a localized component in the Cu diffusion, while the tetrahedral to tetrahedral Cu hops, which contribute to long-range diffusion, are slower. To identify the Cu hoping pathways, we have computed the probability density at 700 K for Cu atoms in (0 0 ½) and (0 0 ¼) planes, which contain the octahedral and tetrahedral Cu sites, respectively (**figure S6** (Supplementary Material[43])). From the probability density plots, it appears that Cu-hopping from a tetrahedral site to another site via octahedral sites are the main hoping channels.

In the superionic hexagonal phase of $Cu_2S$, we found that Cu diffusion is nearly isotropic (**figure S7,** (Supplementary Material[43])), which is in contrast with previously reported[36] AIMD simulations. In the hexagonal phase, to visualize the nature of diffusion, we also plotted the trajectories of a few Cu atoms at 500 K (**figure 6**). Further, by observing the individual Cu trajectories, we can see that Cu also rattles around the 4*g* and 6*b* sites (localized diffusion) followed by a jump to 2*b* sites (long-range diffusion). The superionic-cubic phase of $Cu_2S$ exhibits similar attributes of Cu diffusion as observed in superionic cubic $Cu_2Se$. From the trajectory analysis at 750 K shown in **figure 6**, we found that there are two types of jump present in the system; the first is dwelling between 192*l* and 8*c* sites that contribute to localized diffusion, while the second kind of jump (i.e. tetrahedral to tetrahedral jump) contributes to long-range diffusion.

C. **Dynamic Structure factor** $S(Q, E)$

Several QENS experiments have been reported[21, 23, 42], which addressed the Cu diffusion behavior in superionic cubic phase of $Cu_2Se$. The results have been analyzed using the C-E jump-diffusion model. The first QENS measurements on the superionic phase of $Cu_2Se$ were performed by Dailkin *et al*[23]. The authors used a single Lorentzian in their analysis, equivalent to considering only a single kind of Cu hoping, and estimated $D \sim 6.1 \times 10^{-5}$ cm$^2$s$^{-1}$. Another QENS investigation by Voneshen *et al*[21] showed two different jump time-scales for the localized and long-ranged diffusion, respectively, and estimated a $D \sim 5\text{-}35 \times 10^{-7}$ cm$^2$s$^{-1}$ over 500-900 K. Another recent QENS study at 675 K by Islam *et al*[42] estimated a $D \sim 3.4 \times 10^{-5}$ cm$^2$ s$^{-1}$.

In all the above QENS studies, the analysis was based on the C-E model assuming incoherent scattering due to independent jumps. Here, we have simulated the incoherent dynamical structure factor of Cu in $Cu_2X$ (X=S, Se), $S_{inc}^{Cu}(Q,E)$, at different temperatures ranging from 500K to 900K, to investigate the jump-length distribution and associated timescale and diffusion constants.

***$Cu_2Se$***: Our calculated incoherent dynamical structure factor for Cu in superionic $Cu_2Se$ at 700 K is shown in **figure 7**. To estimate $D$, we use $S_{inc}^{Cu}(Q,E)$, which represents the random/stochastic diffusion process, and justifies the use of the C-E model. We have first analyzed the calculated $S_{inc}^{Cu}(Q,E)$ with a single Lorentzian and noticed a significant discrepancy between the calculated $S_{inc}^{Cu}(Q,E)$ and the fit (**figure S8(a)** (Supplementary Material[43])). Interestingly, two Lorentzian excellently describe the calculated $S_{inc}^{Cu}(Q,E)$, suggesting two kinds of diffusive dynamics present in the system (**figure S8(b)** (Supplementary Material[43])). The widths of the two Lorentzians differ by an order of magnitude, which implies the two dynamics occur on very different time-scales. The fast diffusion dynamics are reflected in the Lorentzian with a larger width, while the slower diffusion gives the smaller Lorentzian width. The $Q$-dependence of the HWHM of the Lorentzian peak provides direct information about diffusion characteristics. The estimated $Q$-dependence of the HWHM of the Lorentzian for superionic cubic $Cu_2Se$ at 500 K and 700 K are shown in **figure 7(b-c).** We find that HWHM of the slower dynamics Lorentzian, $\Gamma_1$ shows a $Q^2$ behavior at low $Q$. This indicates the presence of long-range diffusion. In contrast, the higher value of HWHM from the second Lorentzian, $\Gamma_2$ remains nearly independent of $Q$, signifying the presence of fast localized diffusion, attributed to hopping among 32$f$ sites around the tetrahedral 8$c$ sites.

Our estimated $Q$-dependent $\Gamma_1$ from $S_{inc}(Q, E)$ for the cubic phase of $Cu_2Se$ at 700 K is underestimated by about a factor of 2 compared to the values reported from the QENS measurement at 700 K by Islam *et al*[42]. The $Q$-dependent $\Gamma_1$ at 500 K and 700 K is analyzed with the C-E model. The fitted residence time ***τ***, jump-length *d* and estimated diffusion constant *D* are summarized in **Table II**. Our estimated *D* from the C-E model at 700 K is in fair agreement with the QENS estimated $D \sim 3\times 10^{-5}$ cm$^2$/sec by Islam *et al* [42].

***$Cu_2S$:*** In **figure (7d)**, we show the computed $S_{inc}^{Cu}(Q,E)$ of superionic cubic $Cu_2S$ at 750 K. Here also, two Lorentzian are required to describe the $S_{inc}^{Cu}(Q,E)$ in both the superionic phases *i.e.*, cubic and hexagonal (**figure S8** (Supplementary Material[43])), corresponding to the faster and slower dynamics, respectively. Further, we have used $Q$-dependence of narrower HWHM from $S_{inc}^{Cu}(Q,E)$ to estimate $D$ via a C-E jump diffusion model (**figure 7e**). The fitted parameters and estimated $D$ for superionic hexagonal and cubic phases at 500K and 750 K, are summarized in **Table II**. The estimated $D$ from the C-E model for the hexagonal phase at 500 K and cubic phase at 750 K is ~$6.4\times10^{-6}$ cm$^2$s$^{-1}$ and

~$3.0\times 10^{-5}$ cm$^2$s$^{-1}$, respectively. The presence of the fast localized diffusion, attributed to hopping among 192$l$ sites around the tetrahedral 8$c$ sites, is similar to Cu$_2$Se.

### D. Diffusion Coefficient using various approaches

The MSD plots of **figure 4** are used to calculate the temperature-dependent diffusion coefficients. Further, we have calculated the relative standard deviation (RSD) of the estimated diffusion constant (**TABLE III**) using an empirical relation as described in Ref [61]. RSD of diffusion constant depends on the number of total jump events during the simulation time ($N_{eff}$) as given by

$$RSD = 3.43/\sqrt{N_{eff}} + 0.04 \qquad (9)$$
$$N_{eff} = TMSD/d_{jump}^2 \qquad (10)$$

where TMSD is total mean-squared displacement defined as the sum of all Cu-MSD in the simulation cell, and $d_{jump}$ is mean jump length of Cu atom. In the case of the cubic and the hexagonal phases, the mean jump length is taken as ~3.2 Å. The analysis was performed on 30 ps trajectory length.

Further, we used the MSD estimated $D$ to estimate the energy barrier, $E_a$ for Cu migration in Cu$_2$X (X= S/Se) using Arrhenius relation as given by[62]:

$$\ln(D(T)) = \ln(D_0) - E_a/k_B T \qquad (11)$$

$D_0$ is the prefactor factor; $k_B$ is the Boltzmann constant. The estimated value of the activation energy barrier of Cu atoms is 0.17eV, 0.16 eV, and 0.18 eV for Cu$_2$Se (superionic-cubic), Cu$_2$S (superionic-hexagonal), and Cu$_2$S (superionic-cubic), respectively, as shown in **figure 4**. These values are in fair agreement with reported activation energy barrier ($E_a$~0.15 eV for cubic phase of Cu$_2$Se and $E_a$~0.19 eV for hexagonal phase of Cu$_2$S ) from electronic impedance-spectroscopy[63, 64].

In **Table III**, we summarized the calculated $D$ from different numerical techniques and also show the relative error from these methods. We further estimated the ionic conductivity using the Nerst-Einstien relation. We found that the Cu ionic-conductivity at 700 K in Cu$_2$Se from computed $D$ from different schemes lies in the range of 1.5-3.0 $\Omega^{-1}$ cm$^{-1}$ , which is in excellent agreement with the reported conductivity at 673 K[63]. This further confirms that Cu- diffusion is well captured in our simulations.

## V. CONCLUSIONS

We have presented INS measurements on $Cu_{1.95}Se$, and lattice dynamics, and AIMD simulation in $Cu_2X$ (X=S, Se) to investigate the Cu diffusion. We are able to identify the nature of diffusion at low and moderate temperatures, a solid-like diffusion comprising (a) the localized jumps among a small cluster of sites around the tetrahedral site, and (b) long-range diffusion involving jumps to the first and the second nearest neighbour sites. At high temperatures of ~900 K, the diffusion is liquid-like, which is not restricted to jumps among definite crystal sites. We have also calculated the diffusion coefficients using different approaches, namely, the mean-squared displacements, velocity auto-correlation, and the jump-diffusion model of the dynamical structure factor. While the first two approaches produce consistent results, we find that the jump-diffusion model assuming jumps over a fixed distance in the present case of $Cu_2X$ is an oversimplification and does not produce consistent results of the diffusion coefficient. However, the range of the magnitude of the diffusion coefficient and the derived ionic conductivity are in fair agreement with the reported experimental values.

## ACKNOWLEDGEMENTS

The use of ANUPAM super-computing facility at BARC is acknowledged. SLC acknowledges the financial support of the Indian National Science Academy for the INSA Senior Scientist position.

TABLE I. The intersite jump-length Cu-Cu in $Cu_2Se$ superionic cubic phase, $Cu_2S$ superionic hexagonal and $Cu_2S$ superionic cubic.

| Hopping Cu Wyckoff sites | Hopping distance | | Hopping Cu Wyckoff sites | Hopping distance |
|---|---|---|---|---|
| | $Cu_2Se$(Cubic) | $Cu_2S$(Cubic) | | $Cu_2S$(Hexagonal) |
| (T-T) Hopping in <100> directions (8c-8c) | 2.92 | 2.88 | 4f-6g | 1.24 |
| (T-T) Hopping in <110> directions (8c-8c) | 4.13 | 4.07 | 6g-6g | 1.94 |
| (T-O) Hopping in <111> directions (8c-4b) | 2.53 | 2.49 | 4f-4f | 2.34 |
| (O-O) Hopping in <110> directions (4b-4b) | 4.13 | 4.07 | 2b-6g | 2.6 |

TABLE II. The calculated diffusion coefficient $D$, jump length ($d$) and residence time ($\tau$) of Cu in different phases of $Cu_2X$ (X= S, Se) from fits using C-E model.

| | $d$ (Å) | $\tau$ (ps) | $D$ ($10^{-6}$ cm$^2$/sec) |
|---|---|---|---|
| $Cu_2Se$-Cubic (700 K) from incoherent $S(Q,E)$ from simulation | 3.7 ± 0.1 | 7.7 ± 0.2 | 29 ± 2 |
| $Cu_2Se$-Cubic (500 K) from incoherent $S(Q,E)$ from simulation | 3.5 ± 0.2 | 25 ± 2 | 8 ± 2 |
| $Cu_2Se$-Cubic Islam et al[42] (675 K) from total $S(Q,E)$ from QENS | 2.8 ± 0.2 | 3.7 ± 0.12 | 34 |
| $Cu_2S$-Hexagonal (500 K) from incoherent $S(Q,E)$ from simulation | 3.0 ± 0.2 | 23.4 ± 1.7 | 6.4 ± 1.3 |
| $Cu_2S$-Cubic (750 K) from incoherent $S(Q,E)$ from simulation | 4.1 ± 0.2 | 9.5 ± 0.6 | 30 ± 5 |

TABLE III. The calculated diffusion coefficient $D$ of Cu in different phases of Cu$_2$X (X= S, Se) from suits of numerical techniques. The estimated Relative standard deviation (RSD) of diffusion coefficient (D) as described in Ref [61]. The analysis was performed on 30 ps trajectory length.

| $T$(K) | $D$ from MSD ($10^{-6}$ cm$^2$/sec) | $D$ from VACF ($10^{-6}$ cm$^2$/sec) | $D$ from C-E model fitting ($10^{-6}$ cm$^2$/sec) | PRB[20] ($10^{-6}$ cm$^2$/sec) | | Relative standard deviation (RSD) of D (in %) $3.43/\sqrt{N_{eff}} + 0.04$ |
|---|---|---|---|---|---|---|
| | **Cu$_2$Se (superionic-cubic)** | | | MSD | VACF | |
| 500 | 5.4 | 4.4 | 11.1±0.2 | x | x | 41 |
| 700 | 21 | 17 | 29±2 | x | x | 30 |
| 900 | 36 | 36 | x | 34 | 34 | 21 |
| | **Cu$_2$S (superionic-hexagonal)** | | | | | |
| 500 | 5.2 | 4.3 | 6.4±1.3 | x | x | 28 |
| 700 | 16 | 13 | 29±4 | x | x | 23 |
| | **Cu$_2$S (superionic-cubic)** | | | | | |
| 750 | 17 | 22 | 30±5 | x | x | 42 |
| 900 | 29 | 30 | 35±2 | 33 | 35 | 24 |

**Figure 1.** (Color Online) The crystal structure of (a) Cu$_2$Se and (b) Cu$_2$S in the high-temperature cubic phase (*F*m-3m). The Cu-occupancies are: (a) Cu$_2$Se- 8*c* (0.72), 4*b* (0.0), 32*f* (0.07); and (b) Cu$_2$S- 8*c* (0.26), 4*b* (0.09), 192*l* (0.03). The Cu occupancies are shown by partially colored large and small spheres, while fully occupied S/Se are shown by green spheres (see legends).

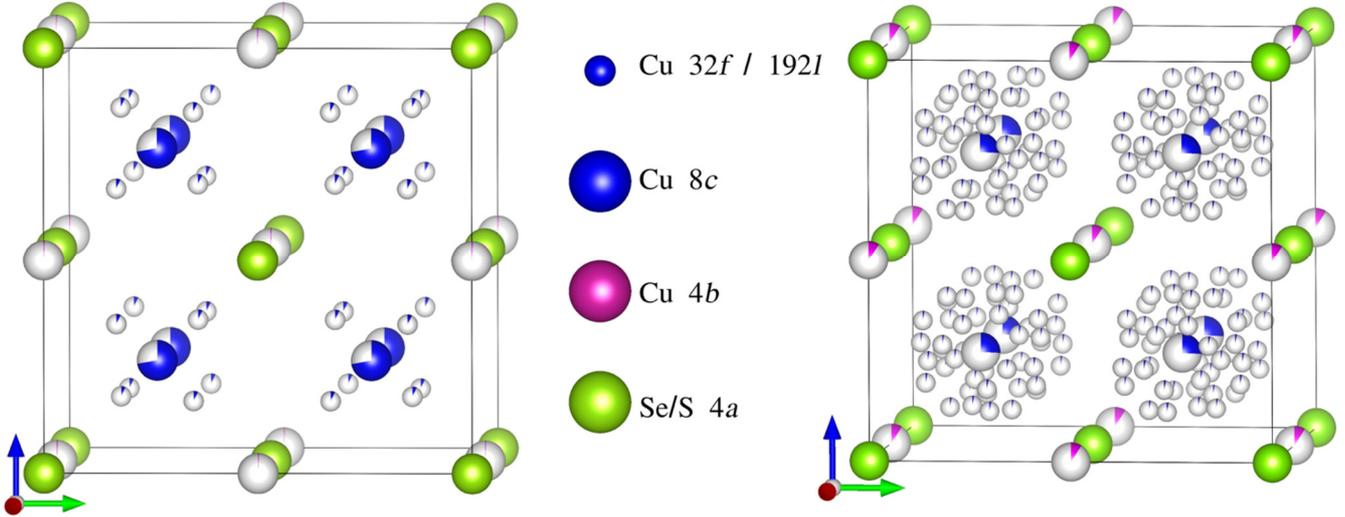

**Figure 2.** (Color Online) (a) Temperature dependence of neutron-weighted PDOS, $g^{(n)}(E)$, of cubic Cu$_{1.95}$Se from INS measurements. (b) AIMD calculated partial contributions of Cu and Se to $g^{(n)}(E)$ and the total at 500 K.

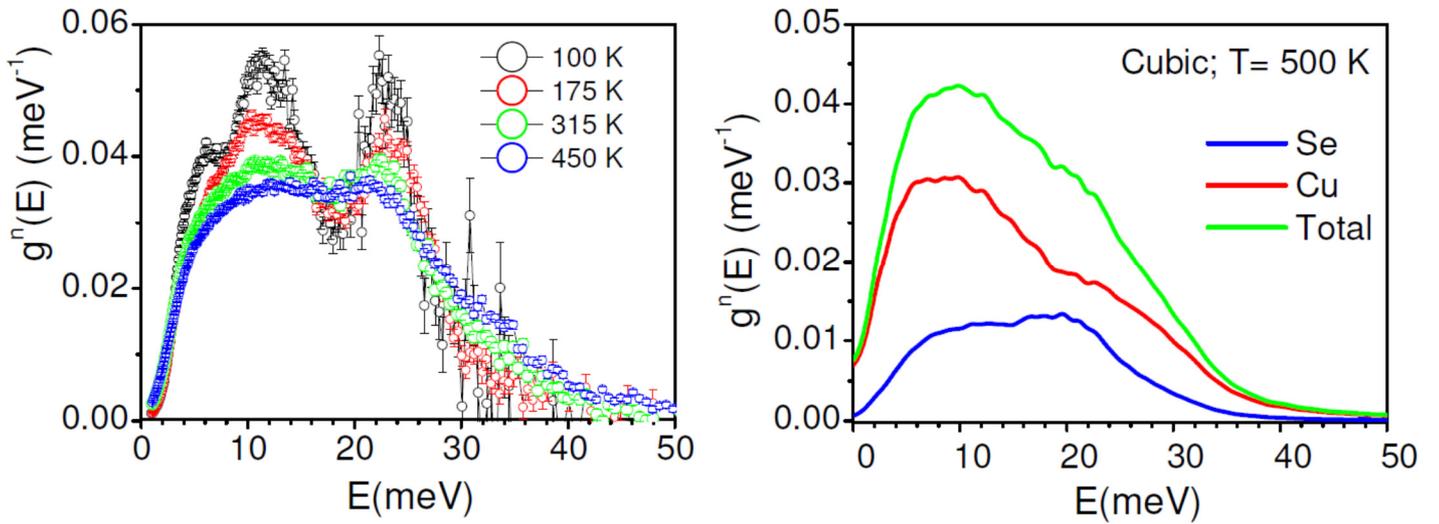

**Figure 3.** (Color Online) The AIMD calculated temperature-dependent $g^{(n)}(E)$ in superionic-cubic phase of **(a)** $Cu_2Se$ and **(b)** $Cu_2S$. The presence of non-zero PDOS at zero energy corresponds to Cu diffusion.

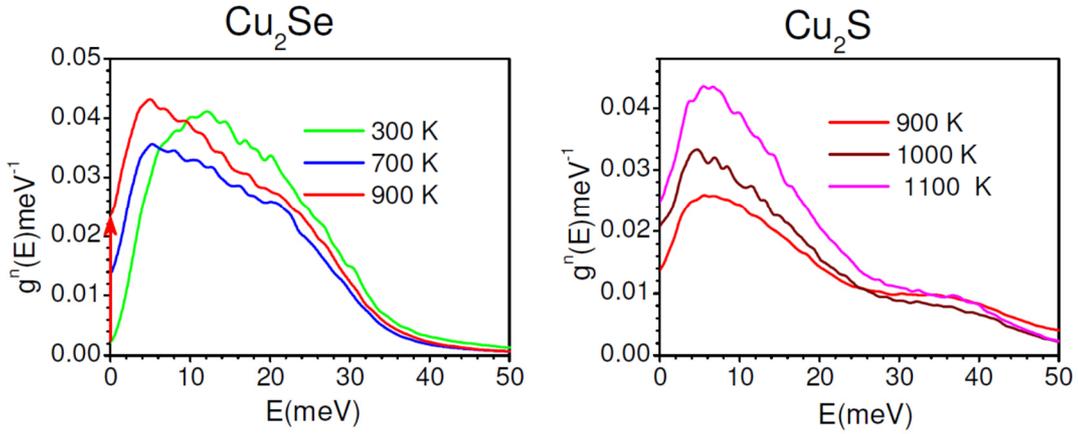

**Figure 4.** (Color Online) (a-c) The calculated MSD of Cu atoms in $Cu_2Se$ and $Cu_2S$ at different temperatures using AIMD simulations performed in superionic-phases. (d-f) Calculated temperature dependence of diffusion coefficient ($D$) estimated from MSD slope and estimated activation barrier $E_a$ for Cu migration.

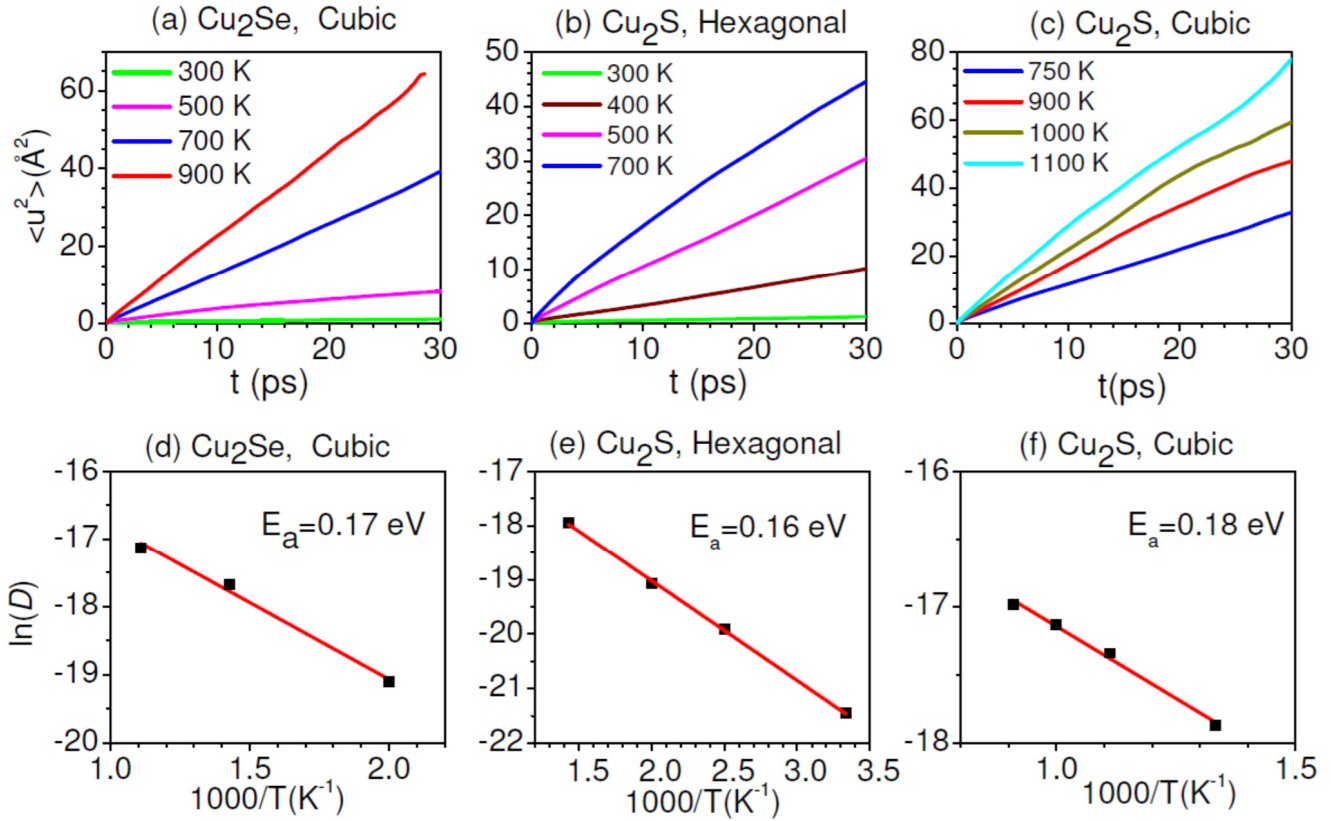

**Figure 5**. (Color Online) The calculated Cu self Van-Hove correlation function $g_s(r,t)$ of (a) $Cu_2Se$ superionic cubic at 500 K, 700 K and 900 K; (b) $Cu_2S$ superionic phases at 500 K (hexagonal), 750 K and 900 K (cubic).

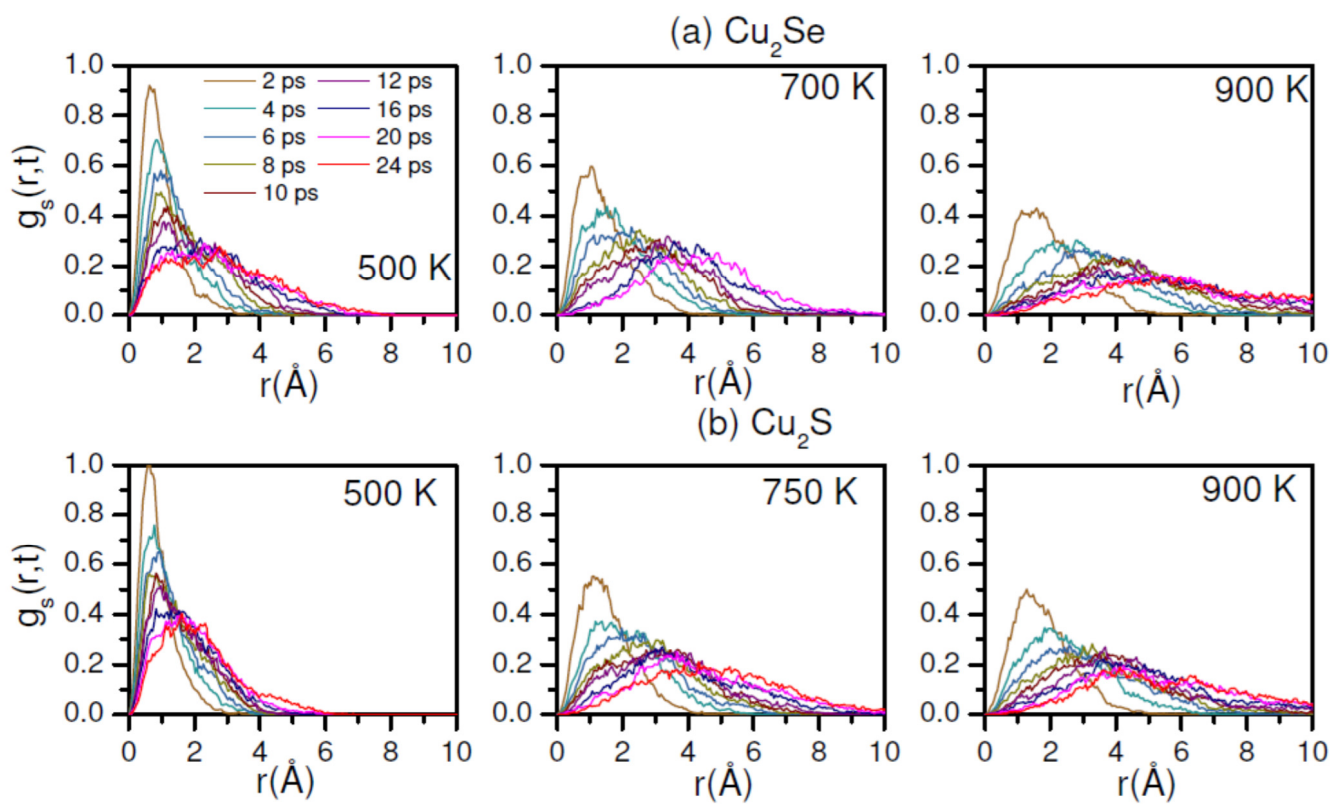

**Figure 6.** The calculated Cu trajectories of a few representative Cu's in superionic phases of $Cu_2X$ (X=S,Se). The Cu trajectories are shown by red dotted line, while the average Cu and X positions are shown by spheres. Details of the average Cu and X sites are shown in legends. Cu trajectory in cubic $Cu_2Se$ (**Top panel**) and cubic $Cu_2S$ (**Middle panel**) shows hops between tetrahedral sites (8$c$) via an octahedral site (4$b$), and also direct jumps. Cu trajectory in hexagonal $Cu_2S$ (**Bottom panel**) shows hops between 2$b$ to 4$f$ sites. In addition, all the trajectories show localized diffusion around the respective sites.

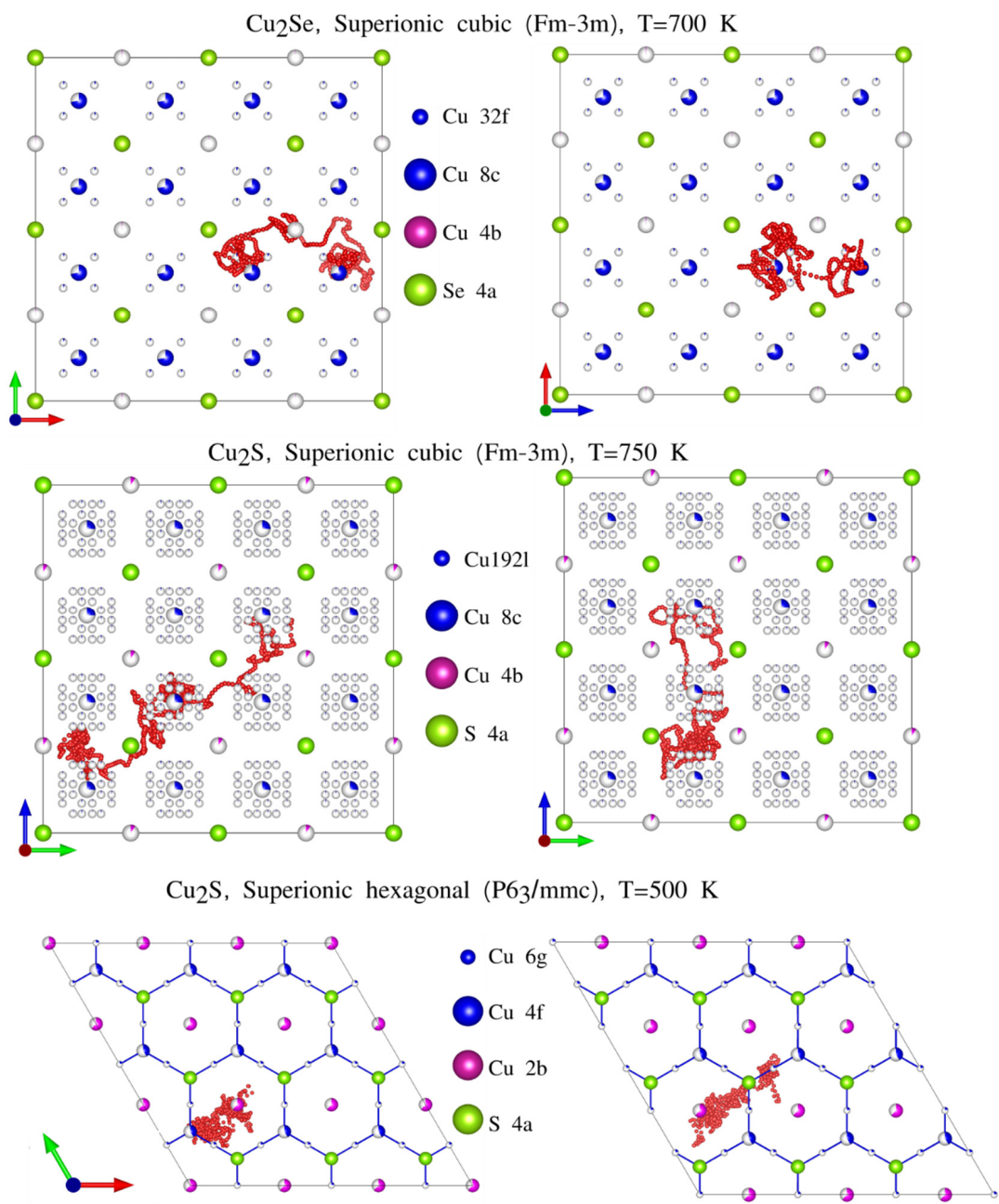

**Figure 7.** (a, d) The calculated incoherent ($S_{inc}^{Cu}(Q,E)$) dynamical structure factors in superionic cubic phases of Cu₂X (X=S, Se). (b, e) The estimated $Q$-dependence of the narrower Lorentzian HWHM from the incoherent $S_{inc}^{Cu}(Q,E)$ in superionic phases of Cu₂Se (Cu₂S) at $T$=500 K (500 K) and 700 K (750 K) are shown by the black and red markers, respectively. The solid lines in panel (b) and (d) are C-E model fit to respective HWHM and (c,f) the estimated $Q$-dependence of the broader Lorentzian HWHM from incoherent $S_{inc}^{Cu}(Q,E)$ in Cu₂Se and Cu₂S.

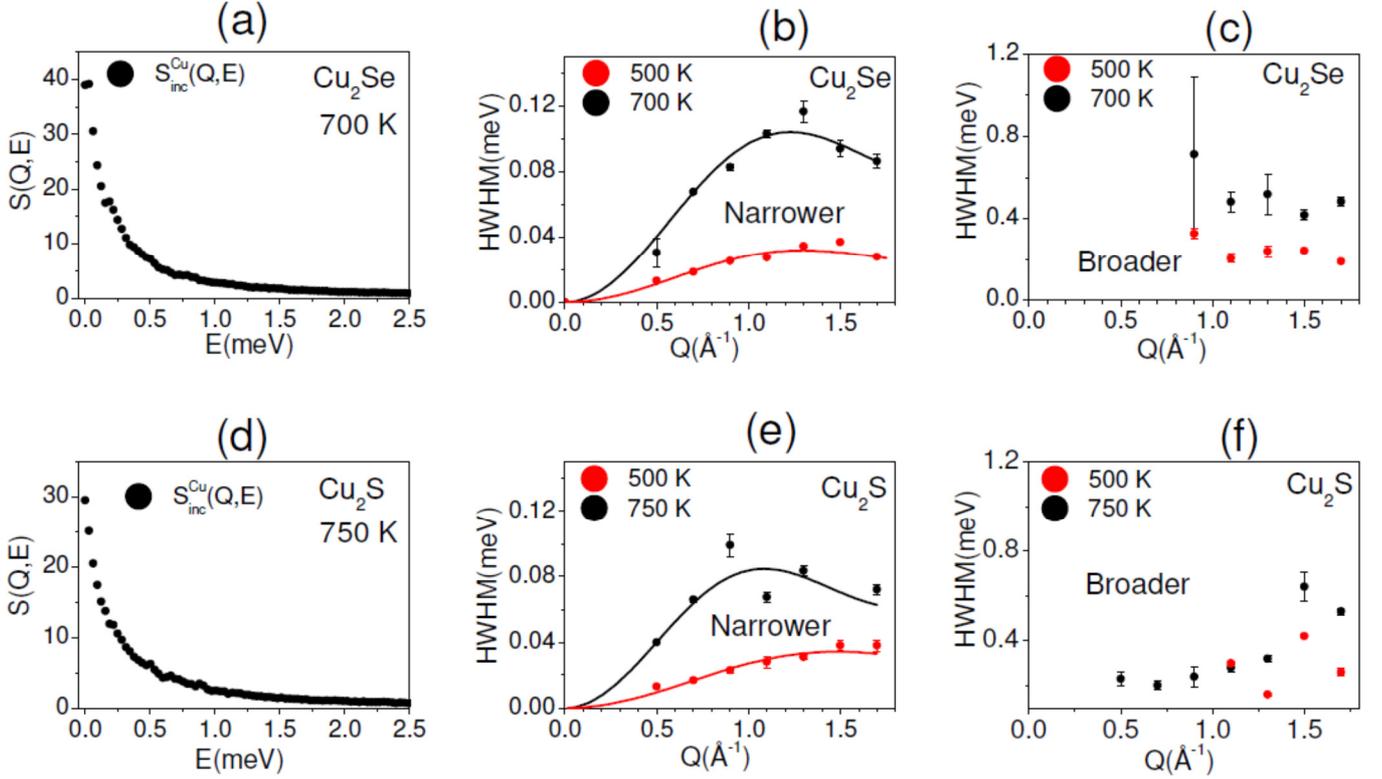

# Supplementary Materials

# Solid-like to Liquid-like Behavior of Cu Diffusion in Superionic Cu$_2$X (X=S, Se): An Inelastic Neutron Scattering and Ab-Initio Molecular Dynamics Investigation

Sajan Kumar[1,2], M. K. Gupta[1,2$], Prabhatasree Goel[1], R. Mittal[1,2*], Olivier Delaire[3], A. Thamizhavel[4], S. Rols[5] and S. L. Chaplot[1,2]

[1]*Solid State Physics Division, Bhabha Atomic Research Centre, Trombay, Mumbai 400085, India*
[2]*Homi Bhabha National Institute, Anushaktinagar, Mumbai 400094, India*
[3]*Department of Mechanical Engineering and Materials Science, Duke University, NC 27708, USA*
[4]*Tata Institute of Fundamental Research, Homi Bhabha Road, Colaba, Mumbai 400005, India*
[5]*Institut Laue-Langevin, BP 156, 38042 Grenoble Cedex 9, France*

Email: rmittal@barc.gov.in[*], mayankg@barc.gov.in[$]

The Cu ionic-conductivity is estimated using the Nerst-Einstien equation

$$\sigma_i = \frac{(Z_i e)^2 C_i D_i}{k_B T}$$

Where $Z_i$ is the charge state of Cu ion, $C_i$ is the concentration of charge particle per cm$^3$, $D_i$ is the diffusion coefficient, k$_B$ is the Boltzmann constant and T is the temperature.

TABLE SI: The structure details of different phases of $Cu_2X$ (X=S, Se) and their temperature range of stability.

| Compound | Phase | T(K) | Space group | Ionic property |
|---|---|---|---|---|
| $Cu_2Se$ | α (Cubic) | >414K | *Fm-3m* | Superionic(disordered) |
|  | β (Monoclinic) | <414K | *C2/c* | non superionic(ordered) |
| $Cu_2S$ | α (Cubic) | ≥ 700 | *Fm-3m* | Superionic(disordered) |
|  | β (Hexagonal) | 370-700K | *P6_3/mmc* | Superionic(disordered) |
|  | Orthorhombic | < 202K | *Pna2_1* | Non Superionic(ordered) |
|  | Tetragonal | < 370 | *P4_32_12* | Non Superionic(ordered) |

TABLE SII: The AIMD calculation details.

| Compound | Phase | Lattice constants(Å) | No. of formula unit/unit cell | Cell size | No. of atoms |
|---|---|---|---|---|---|
| $Cu_2Se$ | α (Cubic) | a=b=c=5.85 ( T= 415 K ) | 4 | 2×2×2 | 96 |
|  | β (Monoclinic) | a=7.14, b=12.4, c=27.4 ( T= 300 K) | 48 | 1×1×1 | 144 |
| $Cu_2S$ | α (Cubic) | a=b=c=5.76 (T = 900 K) | 4 | 2×2×2 | 96 |
|  | β (Hexagonal) | a=3.89, c=6.68 (T 385 K) | 2 | 3×3×2 | 108 |
|  | Orthorhombic | a=5.39, b=5.80, c=5.70 (T= 100K) | 4 | 2×2×2 | 96 |
|  | Tetragonal | a=4.0, c=11.27 (T= 293 K) | 4 | 3×3×1 | 108 |

**Figure S1.** (Color Online)   (a) The measured X-ray diifraction pattern of $Cu_{1.85}Se$ at 300 K, and fitted structure model (Fm-3m).

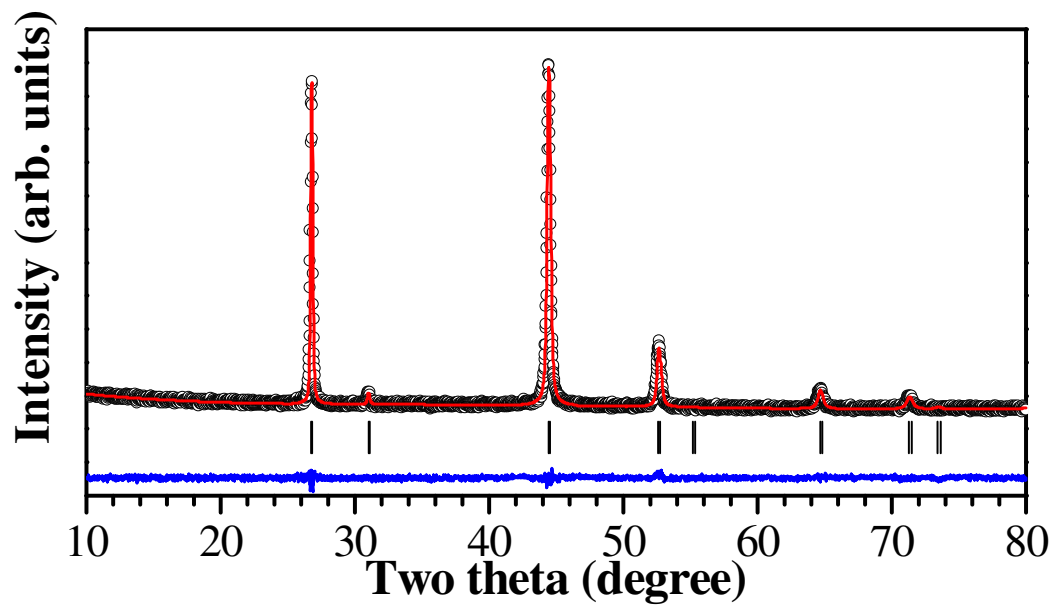

**Figure S2.** (Color Online)  (a) The calculated PDOS at 0 K using harmonic lattice-dynamics (in (a) of monoclinic $Cu_2Se$, (b) orthorhombic $Cu_2S$ and (c) tetragonal $Cu_2S$ phases.

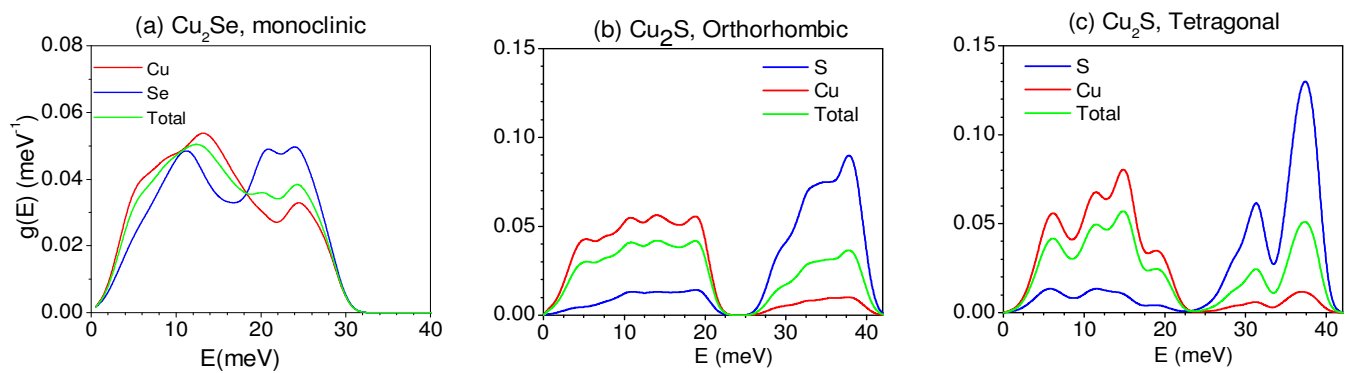

**Figure S3.** (Color Online) The calculated partial contributions of Cu and X (= Se/S) to total PDOS in superionic phase of $Cu_2X$ using AIMD simulations at different temperatures.

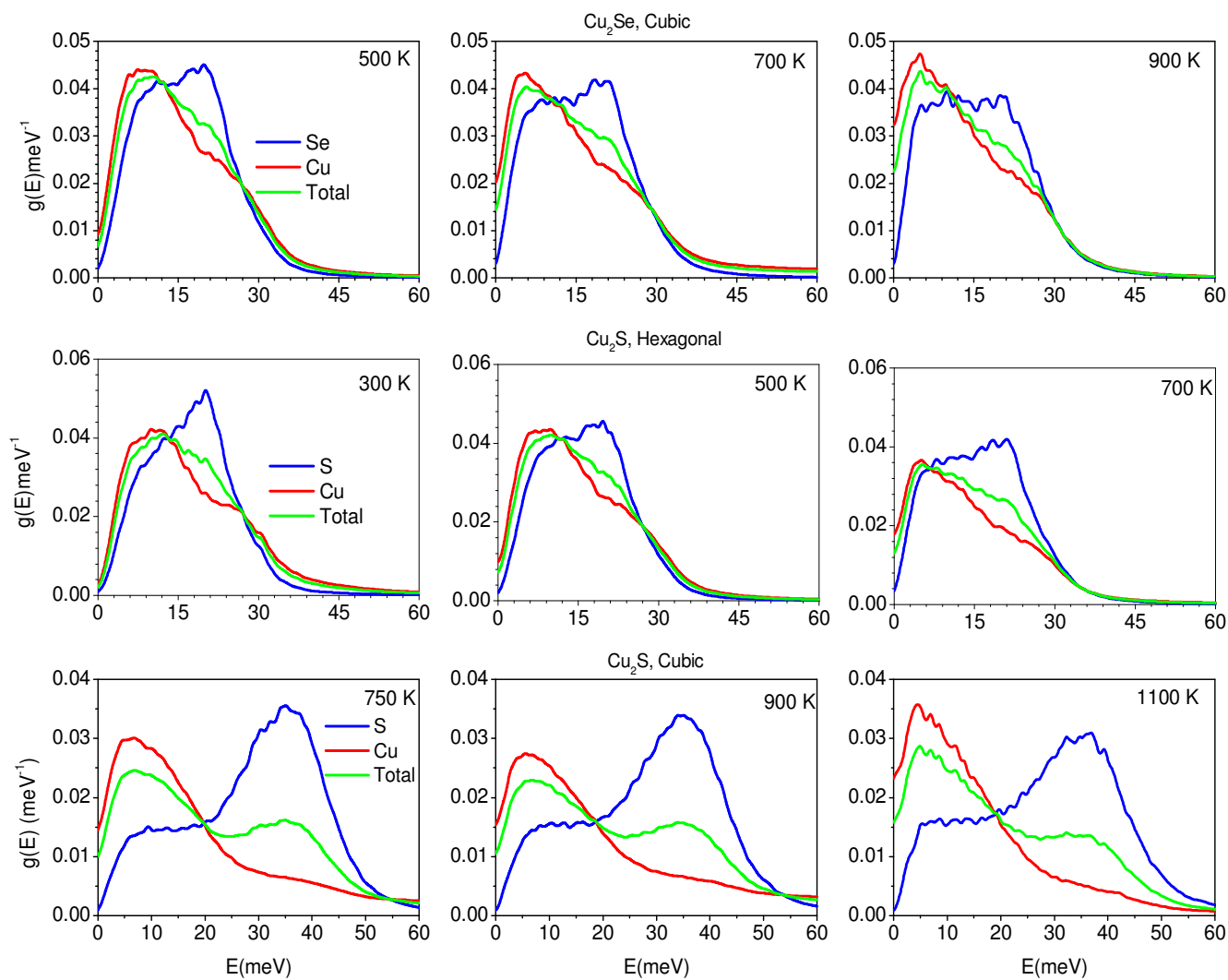

**Figure S4**. (Color Online) The calculated MSD using AIMD trajectories of Cu and X atoms in the superionic phases of $Cu_2X$ (X=S, Se).

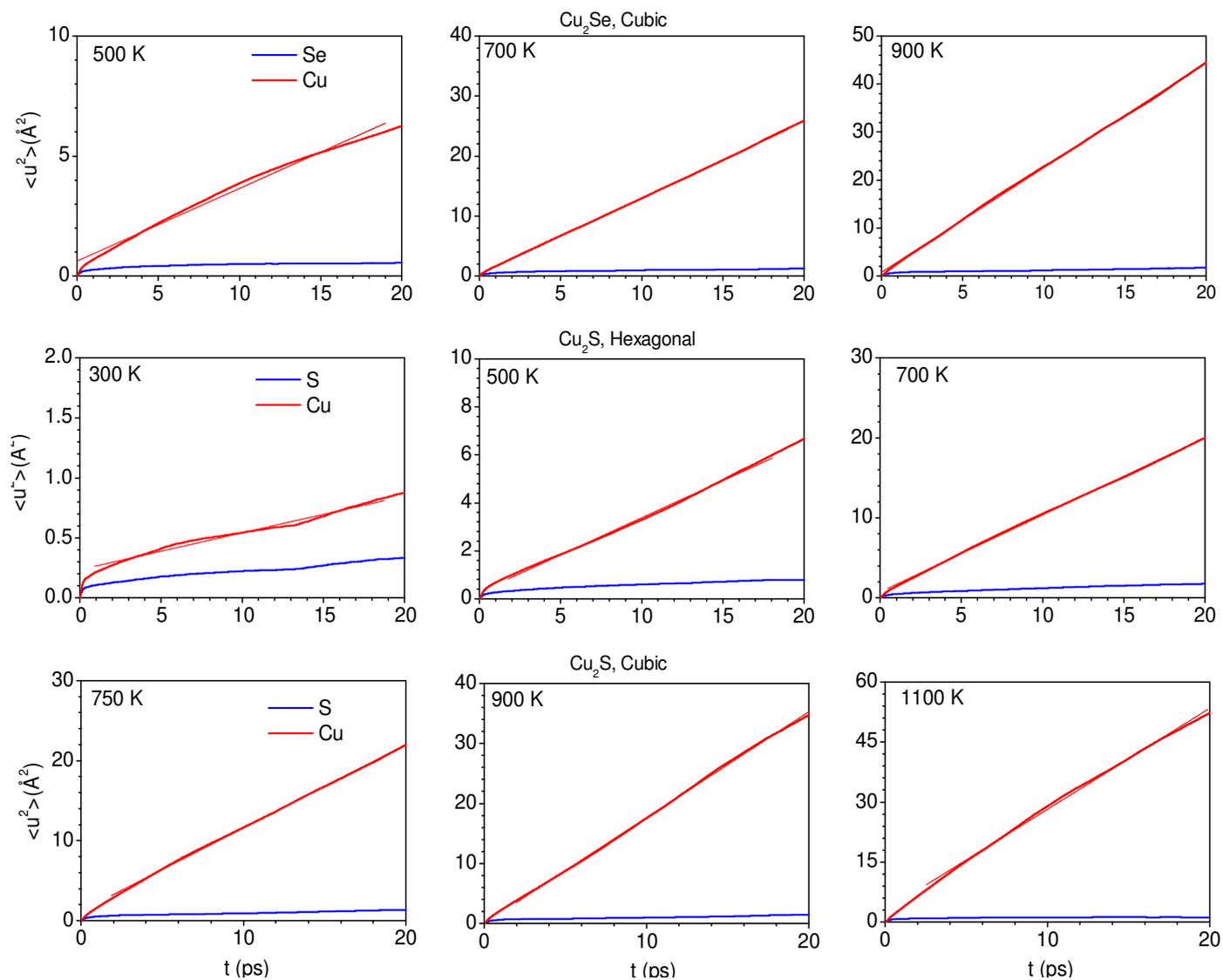

**Figure S5.** (Color Online) The calculated pair distribution function of various pairs of atoms (S-S, Se-Se, Cu-S, Cu-Se, Cu-Cu, and total) in the superionic phases of $Cu_2S$ (hexagonal 300 K, cubic 750 K) and $Cu_2Se$ (cubic 500 K) using AIMD.

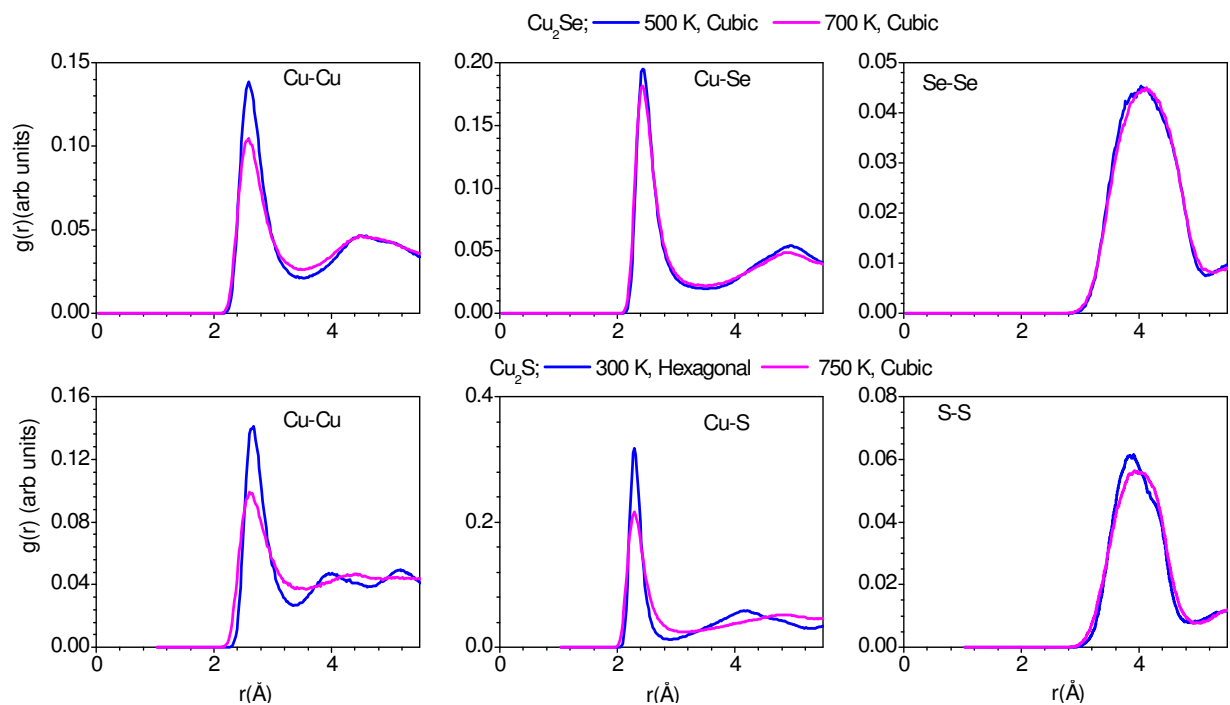

**Figure S6.** (Color Online) (top-panel) The calculated Cu probability map in (0 0 1/2) and (0 0 1/4) plane contains the octahedral sites (pink spheres, Left panels) and tetrahedral sites (blue spheres, Right panels), respectively in cubic phase of $Cu_2Se$ and $Cu_2S$ at 700K and 750 K respectively. The red and blue colors corresponds to the highest and lowest probability.

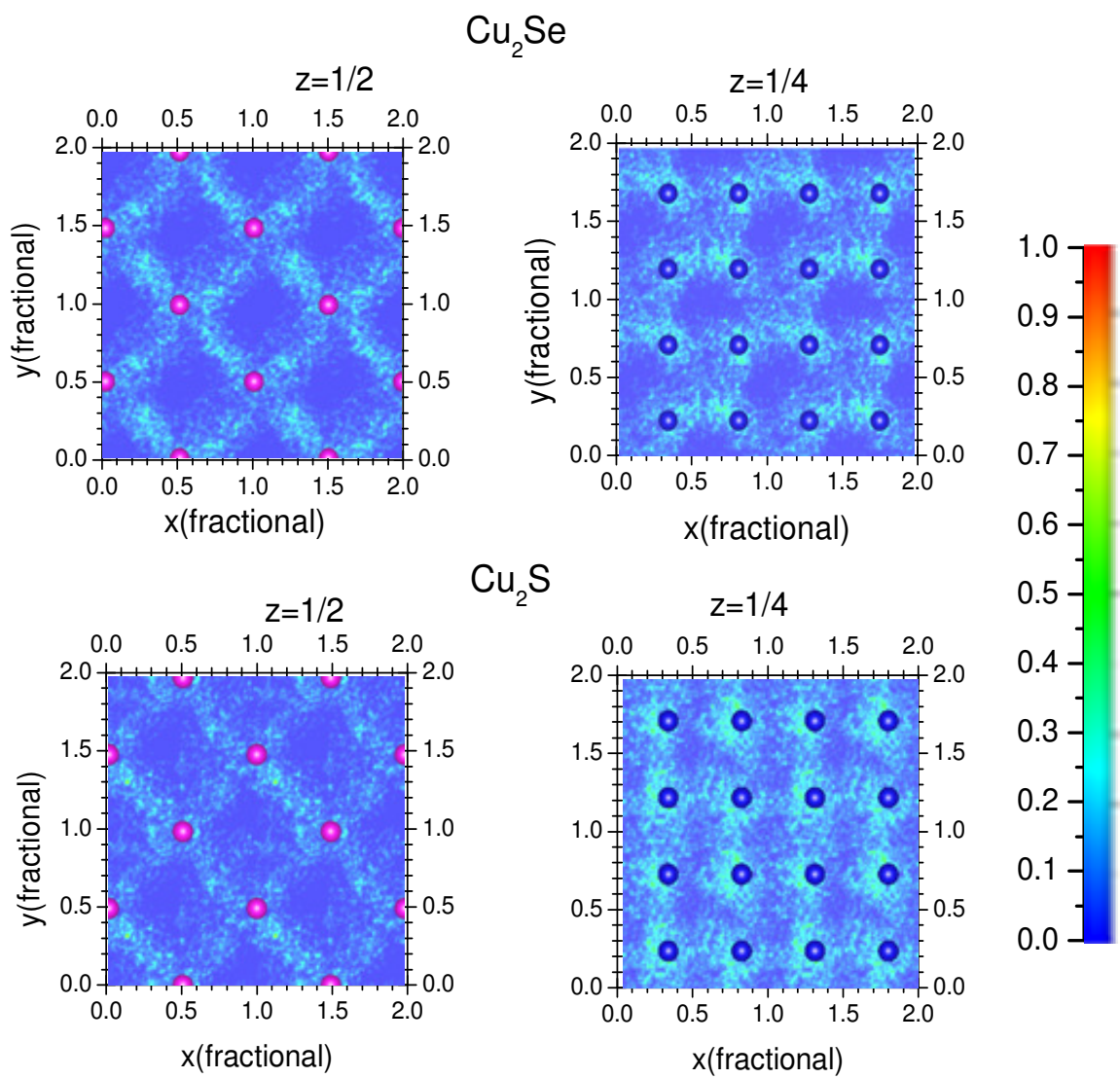

**Figure S7**. (Color Online) (a) The calculated MSD component along Cartesian direction in hexagonal phase of Cu$_2$S at 500 K. The MSD values are nearly isotropic.

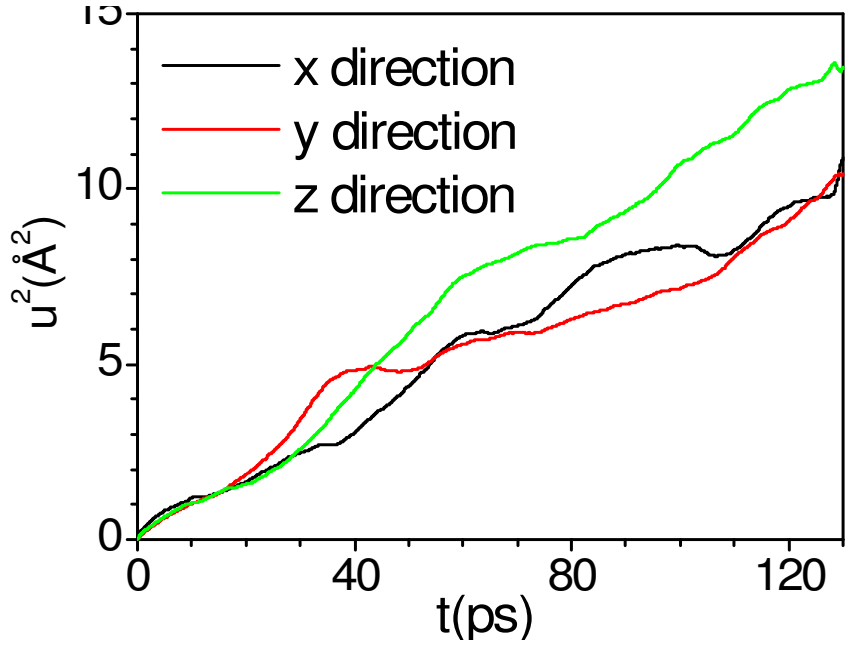

**Figure S8**. (Color Online) (a) The calculated $S_{inc}^{Cu}(Q,E)$ of Cu$_2$Se at 700 K (black dots) analyzed using (a) one Lorentzian and (b) two Lorentzian ($\Gamma_1$, $\Gamma_2$) functions (solid-lines). The two Lorentzian model better describes the $S_{inc}^{Cu}(Q,E)$.

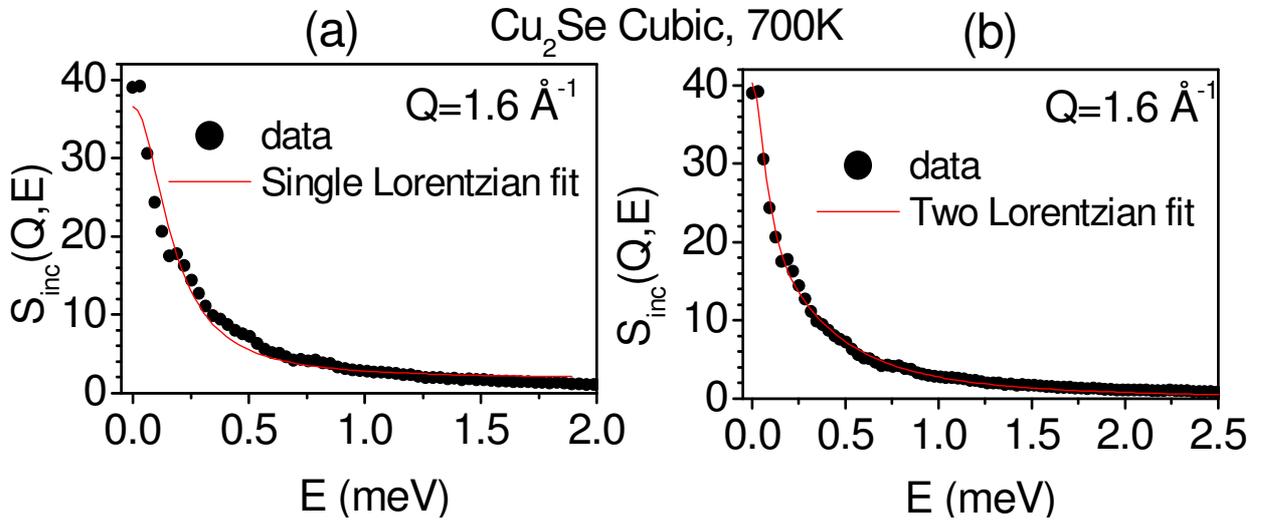